\RequirePackage{snapshot}
\RequirePackage[final]{graphicx}
\documentclass[twocolumn,final,aps,nofootinbib,nobibnotes]{revtex4-2}

\usepackage{amsmath,amssymb,amsfonts,amsthm}

\usepackage{enumerate}
\usepackage{enumitem}
\usepackage{epsfig}
\usepackage{latexsym}
\usepackage{xcolor}
\usepackage[colorinlistoftodos, shadow, textsize=small, obeyDraft]{todonotes}

\usepackage[colorlinks=true,citecolor=blue,linkcolor=blue,final]{hyperref}

\usepackage{comment}

\newcommand{\bite}{\begin{itemize}}
	\newcommand{\eat}{\end{itemize}}
\newcommand{\beq}{\begin{equation}}
	\newcommand{\eeq}{\end{equation}}

\newcommand{\beqa}{\begin{align}}
	\newcommand{\eeqa}{\end{align}}
\newcommand{\barr}{\begin{array}}
	\newcommand{\earr}{\end{array}}

\newcommand{\C}{\mathbb{C}}

\newcommand{\ie}{\textit{i.e.}~}

\newcommand{\bz}{\mathbf{z}}

\newcommand{\mc}[1]{\mathcal{#1}}
\newcommand{\mbb}[1]{\mathbb{#1}}
\newcommand{\mf}[1]{\mathfrak{#1}}

\usepackage{bbold}

\newcommand{\id}{\mathbb{1}}

\newcommand{\vect}[1]{\boldsymbol{#1}}
\newcommand{\bvec}[1]{\boldsymbol{\vec #1}}
\newcommand{\expect}[1]{\langle #1\rangle}
\newcommand{\innerp}[2]{\langle #1 \vert #2 \rangle}
\newcommand{\expectop}[3]{\langle #1 \vert #2 \vert #3 \rangle}
\newcommand{\bra}[1]{\langle #1 \vert}
\newcommand{\ket}[1]{\vert #1 \rangle}

\newcommand{\sltwoc}{\mathfrak{sl}(2,\mathbb{C})}

\newcommand{\norm}[1]{\lVert #1 \rVert}

\newcommand{\rket}[1]{\vert #1 ]}
\newcommand{\rbra}[1]{[ #1 \vert}

\newcommand{\rinnerp}[2]{[ #1 \vert #2 ]}

\newcommand{\bket}[1]{\vert #1 )}

\newcommand{\innerpB}[2]{[ #1 \vert #2 \rangle}

\newcommand{\onehalf}{\frac{1}{2}}

\graphicspath{{figures/}}

\begin{document}

\preprint{This line only printed with preprint option}

\title{Coherent States and Particle Scattering in Loop Quantum Gravity}

\author{Deepak Vaid}
\email{dvaid79@gmail.com}
\affiliation{National Institute of Technology, Karnataka, India}
\author{Devadharsini Suresh}
\affiliation{Indian Institute of Science Education and Research, Pune, India}

\date{\today}

\begin{abstract}
Quantum field theory provides us with the means to calculate scattering amplitudes. In recent years a dramatic new development has lead to great simplification of such calculations. This is based on the discovery of the``amplituhedron'' in the context of scattering of massless gauge bosons in Yang-Mills theory. One of the main challenges facing Loop Quantum Gravity is the lack of a clear description of particle scattering processes and a connection to flat space QFT. Here we show a correspondence between the space of kinematic data of the scattering $ N $ massless particles and $ U(N) $ coherent states in LQG. This correspondence allows us to provide the outlines of a theory of quantum gravity based upon the dynamics of excitations living on the the positive Grassmannian.
\end{abstract}

\maketitle

\tableofcontents

\listoftodos

\section{Introduction}

There are two primary challenges any contending theory of quantum gravity must provide answers to. These are:

\begin{itemize}
	\item[\textbf{A.}] The existence of a consistent semi-classical limit where one can make contact with the usual predictions of continuum quantum field theory.
	\item[\textbf{B.}] The existence of sufficient number of particle species and interactions in order to be able to embed the Standard Model of particle physics within the theory.
\end{itemize}

Loop quantum gravity (LQG) is one such contender but it is yet to satisfactorily answer both of these challenges. There have been several proposals for the item \textbf{(B)} in the form of so-called ``preon'' or ``helon'' models \cite{Bilson-Thompson2005A-topological,Bilson-Thompson2006Quantum,Wan2007On-Braid,Smolin2007Propagation,Bilson-Thompson2008Particle,Wan2009Effective,Vaid2010Embedding} which attempt to exploit topological degrees of freedom of spin network states in order to construct states which can map to, at least, one generation of Standard Model fermions\footnote{For other works which follow a similar line of thought, see \cite{Gresnigt2017Quantum,Asselmeyer-Maluga2019Braids}. There is also a model of elementary particles built upon the properties of division algebras \cite{Furey2010Unified,Furey2014Generations:,Furey2016Standard}, but it is not yet clear as to how this might be related to the braid models of LQG.}.

Even if one is able to satisfactorily embed the Standard Model into LQG by extending the state space to include braid excitations (or perhaps in some other form), the question \textbf{(A)} of making contact with the conventional QFT picture where one could, for instance, describe the scattering of particles around a flat Minkowski background in terms of a S-matrix, still remains out of reach. Though there have been some proposals for how one can construct the graviton propagator \cite{Rovelli2005Graviton} and $ n $-point functions \cite{Modesto2005Particle} in the context of a background independent theory of quantum gravity, there is as yet no satisfactory description of scattering processes in LQG.

In this work we want to point out the existence of certain structures in LQG, which exactly parallel the spinor helicity formalism used to describe the scattering of massless gauge bosons in QFT in the ``on-shell'' framework, first pioneered by Parke and Taylor \cite{Parke1986Amplitude} whose discovery of the Parke-Taylor formula led to a revolution in the field of scattering amplitudes. It turns out that the kinematic space for the scattering of $ n $ massless particles in flat spacetime is the Grassmannian $ Gr_{2,n} $ - the space of $ n $ two-planes in $ \C^n $. This can be demonstrated quite easily by using the spinor helicity formalism to describe the states of massless particles and exploiting Lorentz invariance of the background spacetime.

In LQG the physical Hilbert space consists of arbitrary graphs whose edges are labelled by irreps of $ SU(2) $ and vertices are labelled by $ SU(2) $ invariant tensors. Now, a priori, there is no smooth background spacetime around which one could, for instance, calculate scattering amplitudes or define particle momenta. In QFT particles are labelled by representations of Wigner's ``little group''. However, a state of quantum geometry as represented by a spin network does not possess any inherent notion of Lorentz or Poincare invariance which would allow us to single out particular representations which could be identified as particle states. The goal then is to \emph{construct} states of quantum geometry, formed from suitable superpositions of spin network states, which would approximate \emph{as closely as possible} a semiclassical background geometry.

One such construction exploits the usual quantum mechanical notion of coherent states, or more precisely their extension to generalized coherent states, to construct states of quantum geometry which are peaked around classical geometries. LQG is built upon the Hamiltonian (or $ 3+1 $) approach to general relativity where spacetime consists of three-dimensional spatial hypersurfaces which evolve along timelike curves to generate the full four-dimensional spacetime. The classical phase of general relativity consists of a pair $ (h_{ab}(\vec x), \pi^{cd}(\vec{x'})) $, where the configuration variable $ h_{ab}(\vec x) $ is the intrinsic metric of a spatial hypersurface and the momentum variable $ \pi^{cd}(\vec{x'}) $ is constructed out of the extrinsic curvature of the same hypersurface.

A coherent state of quantum geometry therefore corresponds to a superposition of spin network states which encode information about \emph{both} the intrinsic and the extrinsic metric of the classical three-dimensional geometry we wish to approximate with that state. Operators which represent the intrinsic and extrinsic curvature components should be maximally peaked around some classical values for such a state to represent a semiclassical geometry. Such states have been studied by \cite{Freidel2009The-Fine,Freidel2010UN-Coherent} and are known as $ U(n) $ coherent states. They correspond to a graph consisting of a single $ n $-valent vertex (of the form shown in \autoref{fig:polyhedron-spins}). The remarkable fact is that these coherent states are labeled by elements of the \emph{same} space which labels the kinematic space for the scattering of $ n $ massless particles: the Grassmanian $ Gr_{2,n} $.

It is our contention that this cannot merely be an accidental similarity but is suggestive of a deep underlying connection between the theory of scattering amplitudes in QFT on the one hand and that of coherent states of quantum geometry in LQG, on the other. That this similarity is more than a mere accident is also suggested by the correspondence between the twistorial formulation of LQG and the null twistor formalism which forms the basis of the ``amplituhedron'' program \cite{Arkani-Hamed2014The-Amplituhedron} in the field of scattering amplitudes.

If correct, this correspondence could lead to a fruitful interplay between two approaches which are both radical in their own way and which both aim to provide a description of spacetime where unitarity and locality are emergent concepts rather than intrinsic constructs. At the very least, we feel that the existence of such a correspondence deserves to be noted, even if it does not lead to any radical new developments in either field.

This paper is organized as follows. In \autoref{sec:scattering} we describe the spinor helicity formalism in terms of which the calculation of gluon scattering amplitudes simplifies greatly. We show, following an argument by Arkani-Hamed \cite{Arkani-Hamed2019Spacetime}, that states in the kinematic space of n-gluon scattering can be labelled by elements of the Grassmannian $ Gr_{2,n} $. In \autoref{sec:coherent-states} we describe the formalism for construction of $ U(n) $ coherent states which describe semiclassical states of geometry in LQG. These states turn to be labeled by elements of the same mathematical object as in the case of gluon scattering amplitudes: the Grassmannian $ Gr_{2,n} $. Finally in \autoref{sec:discussion} we discuss this correspondence and what it might imply for the validity of LQG as a theory of quantum gravity with a viable semiclassical limit. We end with some speculations on the relation between particle scattering and the $ \textbf{Complexity} = \textbf{Action} $ and $ \textbf{Complexity} = \textbf{Volume} $ conjectures.

\section{Scattering Amplitudes and the Grassmannian}\label{sec:scattering}

The scattering of elementary particles is the bread and butter of high energy physicists. It is the amplitudes associated with these processes which connect the esoteric techniques of quantum field theory (QFT) with the ``real'' world. The standard machinery for calculating scattering amplitudes involves the use of perturbation theory and the use of Feynman diagrams. Despite the tremendous effectiveness of these techniques\footnote{For instance, in the determination of the fine-structure constant $ \alpha = \frac{e^2}{hc} $ to a precision of eight significant digits.} the Feynman diagram method fails miserably when it comes to helping us understand the dynamics of particles and fields interacting via the strong force.

\begin{figure}
	\centering
	\includegraphics[width=0.8\linewidth]{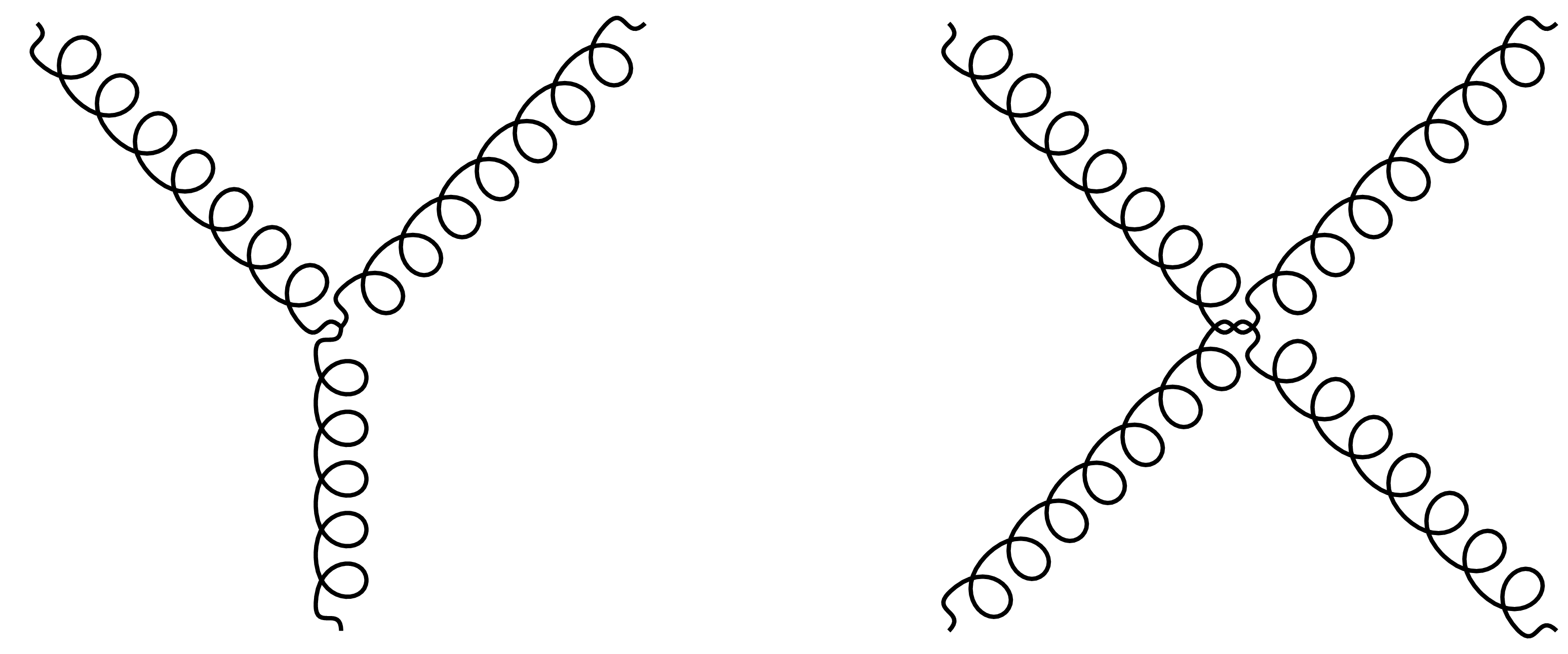}
	\caption{3 and 4 gluon vertices}
	\label{fig:gluon-vertices}
\end{figure}

Quantum chromodynamics (QCD) is the theory of how quarks and gluons - which make up more than 98\% of the ordinary matter content in the Universe - interact with each other to form protons, neutrons and other particles which go by the name of ``hadrons''. QCD can be thought of as a form of quantum electrodynamics (QED), which we understand very well and whose amplitudes can be calculated to great precision with Feynman diagram techniques. However, the essential difference between QCD and QED arises from the fact that in QCD, in contrast to the situation in QED, the force carriers (known as gluons) interact with each other. This leads to the presence of interaction vertices of the form shown \autoref{fig:gluon-vertices} in the calculation of the amplitudes of most QCD processes. The evaluation of such amplitudes involves evaluating Feynman digrams whose number increases exponentially with the total number of particles involved in a given process. However, beginning in the 1980s it began to be noticed that despite the apparently overwhelming complexity of these processes, after summing thousands of Feynman diagrams associated with a given process, the final result would collapse into a single expression of remarkable simplicity. This gave birth to the hope that there were some heretofore unknown symmetries which, if exploited correctly, could lead to dramatic simplifications in such calculations.

The canonical example of such a simplification is in the form of Parke-Taylor formula \cite{Parke1986Amplitude} for the tree-level scattering of $ n $ gluons \footnote{Introductions to the MHV formalism, BCFW recursion relations and related techniques can be found in several references. Some of these include \cite{Dixon1996Calculating,Alday2008Lectures,Feng2012An-Introduction,Elvang2013Scattering,Dixon2013A-brief}}:

\begin{equation}\label{eqn:parke-taylor}
	\mc{A}_n(1^+, \ldots, i^-, \ldots, j^-, \ldots, n^+) = \frac{\innerp{i}{j}^4}{\innerp{1}{2} \innerp{2}{3} \ldots \innerp{n-1}{n} \innerp{n}{1}}.
\end{equation}

for the tree-level amplitude for the scattering of $ n $ gluons. Here the numbers $ 1, \ldots, n $ represent the momenta of the $ n $ gluons written in terms of \emph{spinor helicity} variables and $ \innerp{i}{j} $ is the symplectic inner product of two spinors: $ \innerp{i}{j} = \epsilon^{ab}\lambda_{ia} \lambda_{jb} $, where $ a \in \{1,2\} $ labels the components of each spinor. Now, the simplicity of this expression is striking especially when we consider that for tree-level processes involving $ n > 4 $ gluons, the number of Feynman diagrams which have to be summed over to obtain the final result grows exponentially as a function of $ n $.  For example, we have:
\begin{align*}
    &g+g \longrightarrow g+g \quad &\text{4 diagrams} \\
    &g+g \longrightarrow g+g+g \quad &\text{25 diagrams} \\
    &g+g \longrightarrow g+g+g+g \quad &\text{220 diagrams} 
\end{align*}
and $g+g \longrightarrow 8g$ has more than one million diagrams. This is to be contrasted with the simplicity of \eqref{eqn:parke-taylor} which gives the answer for amplitude for a maximally helicity violating (MHV) tree-level scattering of an arbitrary number of $ n $ gluons.

To proceed further we need to briefly describe the spinor helicity formalism.

\subsection{Spinor Helicity Formalism}\label{sec:spinor-helicity}

It is a well known fact that any vector $ p^\mu := (p^0,\vect{p}) $ in Minkowski spacetime can be mapped to Hermitian matrices as follows:
\begin{equation}\label{eqn:spinor-mapping}
	p_{\alpha \dot \alpha} = \sigma^\mu_{\alpha \dot \alpha} p_\mu
\end{equation}
where $ \sigma^\mu := (\id, \vect{\sigma}) $ are the three Pauli matrices supplemented with the identity matrix. Together, these four matrices provide a basis for the space of all $ 2 \times 2 $ Hermitian matrices. Here $ \alpha, \dot \alpha \in \{1,2\} $ label the matrix components, and $ \mu, \nu, \ldots \in \{0,1,2,3\} $ are spacetime indices. Working in the Weyl basis for the Pauli matrices we find that the momentum matrix can be written as:
\begin{equation}\label{eqn:momentum-matrix}
	p_{\alpha \dot \alpha} := \begin{pmatrix}
		p^0 + p^3 & p^1 + i p^2 \\
		p^1 - i p^2 & p^2 - p^3
		\end{pmatrix}
\end{equation}
It is easy to verify that the determinant of this matrix is equal to the squared norm of the associated 4-vector:
\begin{equation}\label{eqn:mom-determinant}
	\det{p} = p^\mu p_\mu = p^2
\end{equation}
Now, in order to describe the scattering of $ n $-gluons we need to specify $ n $ momentum vectors: $ \{p^1_\mu, p^2_\mu, \ldots, p^n_\mu\} $. Since gluons are massless, their momenta must be null vectors: $ p^2 = 0 $. The corresponding matrices will therefore have vanishing determinant and will be of rank 1. Now any rank 1 matrix can be written as the product of two vectors as follows:
\begin{equation}\label{eqn:null-momenta}
	p_{\alpha \dot \alpha} = \lambda_\alpha \tilde{\lambda}_{\dot \alpha}
\end{equation}
where $ \lambda_\alpha, \tilde \lambda_{\dot \alpha} $ are two 2-component vectors. The requirement that the gluon momentum be real implies that these two vectors are not independent but satisfy: $ \tilde \lambda_{\dot \alpha} = \pm (\lambda^*_\alpha) $ \cite{Arkani-Hamed2017Scattering}.

\subsection{Grassmannian}\label{sec:grassmannian}

Now any such set of $ n $ null momenta obeys a certain symmetry which becomes more apparent when we express the initial data in the following manner:
\begin{equation}\label{eqn:n-vectors}
	\begin{pmatrix}
		\bvec{a} \\
		\bvec{b}
	\end{pmatrix} = \begin{pmatrix}
	\lambda^1_1 & \lambda^2_1 & \ldots & \lambda^n_1 \\
	\lambda^1_2 & \lambda^2_2 & \ldots & \lambda^n_2
\end{pmatrix}
\end{equation}
where we have collected the first (resp. second) component of each of the $ n $ null spinors into an $ n $ dimensional (complex) vector $ \bvec{a} $ (resp. $ \bvec{b} $). Thus the kinematic space for $ n $ gluon scattering can be specified in terms of two $ n $ dimensional (complex) vectors $ \bvec{a}, \bvec{b} \in \C^n  $. However, this specification is only with respect to a given observer. One can always perform a Lorentz transformation to a different frame of reference where the initial data would be represented by a \textit{different} set of $ n $ dim vectors $ \bvec{a'}, \bvec{b'} \in \C^n $.

It turns out that there is a very simple relation between two such sets of vectors representing the kinematic data in two different Lorentz frames. Since Lorentz transformations act as linear transformations given by elements of $ \sltwoc $, each pair of components of the unprimed and primed vectors will be related to each other by some linear transformation, i.e.:
\begin{align}\label{eqn:linear-sl2}
		a'_i & = \alpha a_i + \beta b_i \nonumber \\
		b'_i & = \gamma a_i + \delta b_i
\end{align}
where the coefficients of the above linear transformations are simply the elements of the $ \sltwoc $ matrix generating the Lorentz transformation. Consequently the new pair of vectors $ (\bvec{a'}, \bvec{b'}) $ can be written as a linear combination of the old pair of vectors $ (\bvec{a}, \bvec{b}) $. Now any two (non-collinear) vectors in $ \C^n $ will span a two-dimensional plane in $ \C^n $. The fact that the Lorentz transformed kinematic data are linear combinations of the original vectors implies that under Lorentz transformations the kinematic data associated with a given $ n $-gluon scattering amplitude continues to remain in the \textit{same} two-dimensional plane of $ \C^n $. This leads us to the following conclusion:

\begin{quote}
	\textit{The space of kinematic data for the scattering of $ n $ gluons in Minkowski space consists of the set of all two-planes in $ \C^n $, also known as the Grassmannian $ Gr_{2,n} $.}
\end{quote}

\section{$ U(n) $ Coherent States in LQG}\label{sec:coherent-states}

\begin{figure}
	\centering
	\includegraphics[width=0.7\linewidth]{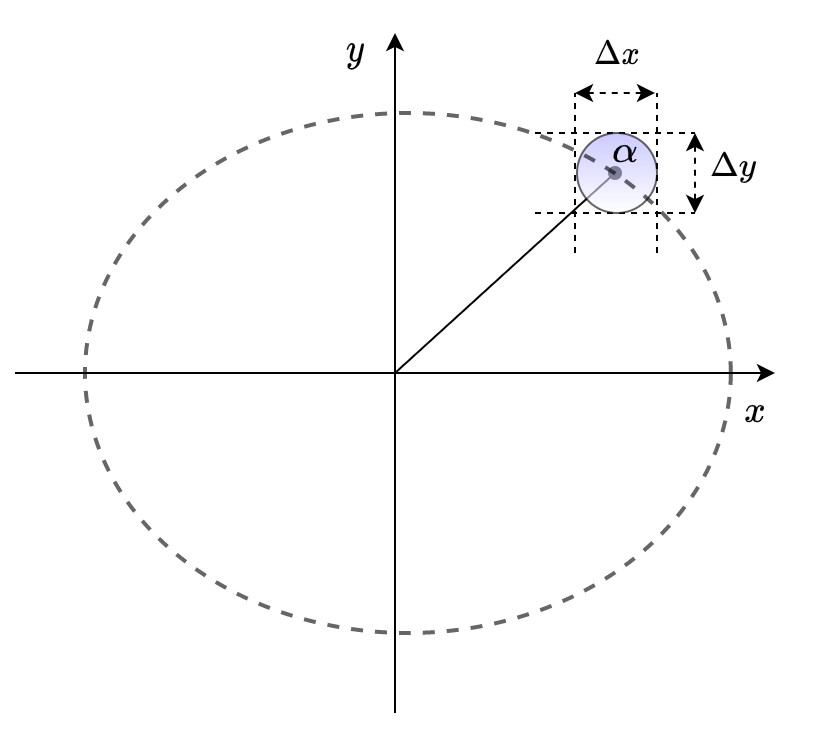}
	\caption{The phase space of a simple harmonic oscillator. The coherent state $ \ket{\alpha} $ saturates the Heisenberg bound $ \Delta x \Delta y = \hbar/2 $. The large ellipse represents the time evolution of $ \ket{\alpha} $ evaluated using Schrodinger's equation and coincides with the trajectory of a classical oscillator. The small circle is the region of uncertainty around the classical phase space point whose co-ordinates are given by $ \expect{\hat x} = \mf{Re}(\alpha),~ \expect{\hat y} = \mf{Im}(\alpha) $.}
	\label{fig:sho-phase-space}
\end{figure}

The question of a consistent semi-classical limit is one that faces \textit{any} quantum theory, not only those of gravity. In ordinary quantum mechanics and also in quantum field theory one of the standard approaches for constructing states which are ``as close as possible'' to states of the classical theory is that of coherent states. A coherent state is one which saturates the Heisenberg uncertainty bound for some pair of non-commuting quantum operators. For instance, for the simple harmonic oscillator states which satisfy this criterion are eigenstates of the annihilation operator:
\begin{equation}\label{eqn:sho-coherent-state}
	a \ket{\alpha} = \alpha \ket{\alpha}
\end{equation}
where $ \alpha \in \C $. The uncertainty in the expectation values of momentum and position operators acting on these states saturates the Heisenberg bound. In particular one has: $ \expect{\hat x} = \mf{Re}(\alpha)$ and $ \expect{\hat p} = \mf{Im}(\alpha) $ and $ \alpha $ therefore corresponds to a point in the phase space of the harmonic oscillator as shown in the \autoref{fig:sho-phase-space}. Our objective, in pursuit of criterion $ \textbf{(A)} $ is to construct semi-classical states of geometry in LQG. One way to do this is by constructing $ U(N) $ coherent states associated with a single vertex of degree $ n $.

\subsection{LQG Phase Space}

In LQG classical geometrical observables such as lengths, areas and volumes are replaced by operators which act on the state space consisting of spin networks. A spin network state is specified in terms of the triple: $ \{\Gamma, \{j_i\}, \{v_k\} \} $, where $ \Gamma $ is a directed graph consisting of $ N_e $ edges and $ N_v $ vertices; $ \{ j_i; i \in 1\ldots N_e \} $ are $ SU(2) $ group representation labels assigned to each edge and $ \{v_k; k \in 1 \ldots N_v \} $ are intertwining operators assigned to each vertex. Area operators act on the edge labels and volume operators act on the intertwiners. With each edge one can associate a quantum of area which is the area of a surface transversally intersected by that edge. With each vertex one can associate a quantum of volume which corresponds to the volume of a polyhedron centered on that vertex and which has as many faces as the degree\footnote{In graph theory, the ``degree'' of a given vertex is the number of edges attached to that vertex} of that vertex.

The Hilbert space consists of cylindrical functions $ \psi(h_1, \ldots, h_{N_e}) $ where $ h_1, \ldots, h_{N_e} $ are $ SU(2) $ valued holonomies of the Ashtekar-Barbero connection along each edge. The (kinematic) Hilbert space of a graph thus consists square integrable functions on $ N_e $ copies of $ SU(2) $:
\begin{equation}\label{eqn:lqg-hilbert-1}
	\mc H_\Gamma = \mc L^2(SU(2)^{N_e})
\end{equation}

\begin{figure}
	\centering
	\includegraphics[width=\linewidth]{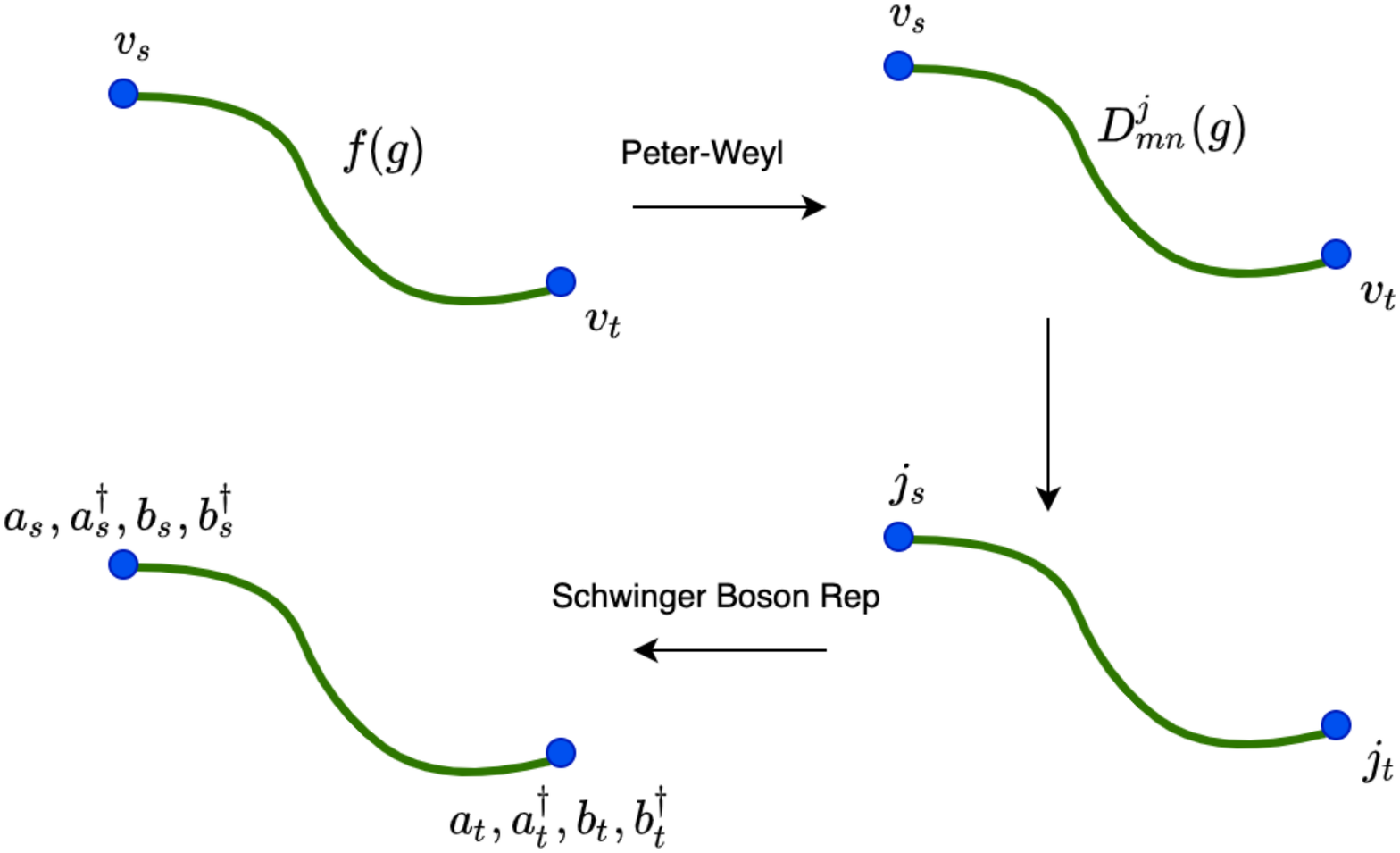}
	\caption{Illustration of steps leading from cylindrical functions of edge holonomies to Schwinger Boson basis with pairs of ladder operators assigned to the source and target vertices respectively.}
	\label{fig:schwinger-boson-rep}
\end{figure}

We can use the Peter-Weyl theorem to write functions on $ SU(2) $ in terms of spin labels (representations of $ SU(2) $) living on each edge:
\begin{equation}\label{eqn:peter-weyl}
	\psi(g) = \sum_{j;mn} f^{mn}_j D^j_{mn}(g)
\end{equation}
where $ \psi(g):SU(2) \rightarrow \C $ is a function of the holonomy, $ j $ labels the different irreps of $ SU(2) $, $ D^j_{mn} $ are the matrix elements of the group element $ g $ in the irrep $ j $ and $ f^{mn}_j $ are the coefficients of the function in the spin basis. Now each matrix $ D^j $ is $ (2j+1) \times (2j+1) $ matrix and therefore can be viewed as a map from one copy of $ \mc{H}_j $ to another copy:
\begin{equation}\label{eqn:irrep-map}
	D^j : \mc{H}_j \rightarrow \mc{H}_j
\end{equation}
Alternatively $ D^j $ can also be viewed as an element of $ \mc{H} \otimes \mc{H} $. Therefore the space of square-integrable functions on $ SU(2) $ can be written in terms of a direct sum of all irreps of two copies of $ SU(2) $:
\begin{equation}\label{eqn:edge-hilbert-space}
	\mc{L}^2(SU(2)) = \oplus_j (\mc{H} \otimes \mc{H})
\end{equation}
This is what leads us to what will be the most crucial point going forwards: \textit{with each edge there are associated two copies of the Hilbert space of a spin $ j $ particle}. This is shown in the upper half \autoref{fig:schwinger-boson-rep}.

\subsection{Area and Volume Operators}\label{sec:geometric-ops}

The spin labels on each edge carry physical information about the geometry represented by the spin network state. Consider an abstract surface pierced by a single edge carrying a spin $ j $. Then from LQG we know that that edge endows the surface with a quantum of area which depends on the spin label. There are two proposals for the area associated with a puncture. These are:
\begin{align}\label{eqn:area-operators}
	A_{RS} & = l_p^2 \sqrt{j(j+1)} \\
	A_{FL} & = l_p^2 j
\end{align}

\begin{figure}
	\centering
	\includegraphics[width=0.5\linewidth]{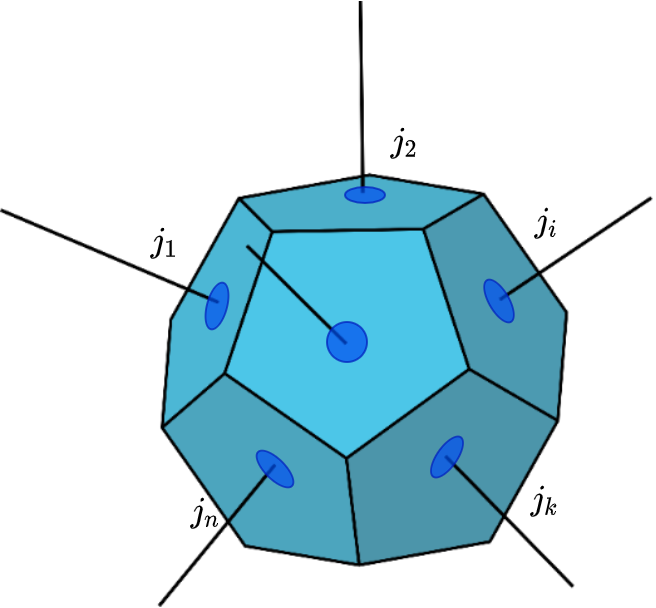}
	\caption{An example of a spin-network vertex, with the associated polyhedron. The faces of the polyhedron are transverse to the edges attached to the vertex. The vertex itself lies in the interior of the polyhedron. Each edge punctures one of the faces and assigns it a quantum of area depending on the spin rep labelling the edge.}
	\label{fig:polyhedron-spins}
\end{figure}

The first is the proposal of Rovelli and Smolin that the area of a puncture is proportional to the Casimir of the spin $ j $. The second is the proposal of Freidel and Livine (and the one we will use in this work) where the area is simply proportional to the highest eigenvalue of the spin $ j $ representation. In this picture we can view any vertex of a spin network as being dual to a polyhedron whose faces are pierced by edges adjacent to that vertex as in \autoref{fig:polyhedron-spins}. In order that the faces associated with each edge, adjacent to a given vertex $ v $, can be glued together consistently to form a polyhedron without any ``gaps'' the following constraint must be satisfied by the spin-network state at each vertex:
\begin{equation}\label{eqn:gauss-constraint}
	(\vec{J_1} + \vec{J_2} + \ldots \vec{J_n})\ket{\Psi_\Gamma} = 0,
\end{equation}
where $ \vec{J_1}, \ldots , \vec{J_n} $ are the angular momentum operators associated with each edge adjacent to the given vertex. This is a statement of the Gauss constraint which says that the (physical) spin network state should be invariant under the action of $ SU(2) $ gauge tranformations acting on its vertices. Now the Hilbert space of all the edges adjacent to a given vertex can be expressed as the tensor product:
\begin{equation}\label{eqn:adjacent-edge-hilbert-space}
	\mc{H}_v = \mc{H}_1 \otimes \mc{H}_2 \otimes \ldots \otimes \mc{H}_n,
\end{equation}
where $n$ is the number of edges attached to the vertex $ v $. Imposing the constraint \eqref{eqn:gauss-constraint} now requires that we assign an invariant tensor to each vertex, \ie an element of the tensor product space $ \mc{H}_v $ which is invariant under $ SU(2) $ rotations generated by \eqref{eqn:gauss-constraint}. On this space of invariant tensors $ \text{Inv}_{SU(2)}(\mc{H}_v) $ one can define a volume operator whose eigenvalues are the possible volumes of the allowed polyhedra dual to that vertex.

\subsection{Schwinger Boson Representation of Edge Hilbert Space}

We now exploit the fact \cite{Girelli2005Reconstructing,Livine2006Reconstructing} that one can express the Hilbert space of a \textit{single} spin $ j $ particle in terms of \textit{two} simple harmonic oscillators as follows. The $ \mf{su}(2) $ algebra is spanned by the operators $ J_x, J_y, J_z $. We introduce two sets of ladder operators $ a, a^\dag, b, b^\dag $ in terms of which the $ \mf{su}(2) $ operators can be expressed as follows:
\begin{equation}\label{eqn:schwinger-boson-rep}
	J_z = \onehalf (a^\dag a - b^\dag b); \quad J_+ = a^\dag b; \quad J_- = b^\dag a
\end{equation}
where $ J_{\pm} = J_x \pm i J_y $ are the usual angular momentum raising and lowering operators and $ a,a^\dag, b, b^\dag $ satisfy the familiar harmonic oscillator commutation relations:
\begin{equation}\label{eqn:sho-commutator}
	[a,a^\dag] = [b, b^\dag] = 1; \quad [a,b] =  [a, b^\dag] = 0
\end{equation}
It is easy to check that the operators \eqref{eqn:schwinger-boson-rep} defined in this way satisfy the usual $ \mf{su}(2) $ algebra: $ [J_i, J_j] = i \epsilon_{ijk} J_k $. This procedure is illustrated in the lower half of \autoref{fig:schwinger-boson-rep}.

\begin{figure}
	\centering
	\includegraphics[width=0.5\linewidth]{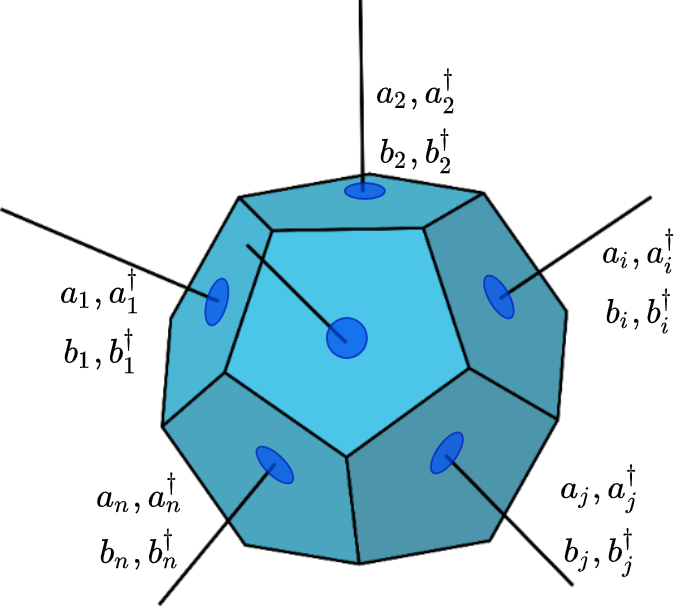}
	\caption{Applying the Schwinger boson trick to each spin label on all the edges.}
	\label{fig:polyhedron-bosons}
\end{figure}

States in the Schwinger boson representation are given by acting on the harmonic oscillator Fock vacuum with the number operators for the two oscillators:
\begin{equation}\label{eqn:schwinger-states}
	\ket{n_a, n_b} = \frac{(a^\dag)^{n_a} (b^\dag)^{n_b}}{\sqrt{n_a ! n_b !}}\ket{0,0}
\end{equation}
The operator for the total energy of such a state is given by:
\begin{equation}\label{eqn:schwinger-energy-op}
	\mc{E} = \onehalf(a^\dag a + b^\dag b),
\end{equation}
with eigenvalues:
\begin{equation}\label{eqn:schwinger-energy-eigenvalues}
	\mc{E} \ket{n_a, n_b} = \onehalf (n_a + n_b) \ket{n_a, n_b}
\end{equation}
Similarly the eigenvalues of the $ J_z $ operator defined in \eqref{eqn:schwinger-boson-rep} are given by:
\begin{equation}\label{eqn:j-z-defn}
	J_z \ket{n_a, n_b} = \onehalf (n_a - n_b) \ket{n_a, n_b}
\end{equation}
This tells that we can identify $ n_a - n_b $ as the magnetic quantum number $ m $ for an angular momentum state $ \ket{j;m} $. Now by explicit calculation using \eqref{eqn:schwinger-boson-rep} and the definition of the energy operator \eqref{eqn:schwinger-energy-op} one can verify that:
\begin{equation}\label{eqn:j-square-defn}
	J^2 = \mc{E}(\mc{E}+1),
\end{equation}
where $ J^2 = J_x^2 + J_y^2 + J_z^2 $ is the $ \mf{su}(2) $ Casimir with eigenvalue $ j(j+1) $ when acting on the state $ \ket{j;m} $. This tells us that we can identify the eigenvalue of the energy operator $ \mc{E} $ with the angular momentum quantum number $ j $. So finally we see that the states of the harmonic oscillator space are in one-to-one correspondence with the usual angular momentum eigenbasis:
\begin{equation}\label{eqn:state-correspondence}
	\ket{j;m} \equiv \ket{n_a, n_b}; \quad j = \onehalf(n_a + n_b);~ m = \onehalf(n_a - n_b),
\end{equation}
where $ j $  is a measure of the total area associated with the given edge with the precise expression depending on which of the prescriptions in \eqref{eqn:area-operators} are used for the area operator.

\subsection{$ \mf{u}(n) $ Lie Algebra}\label{sec:u-n-algebra}

Now, given a vertex $ v $, we repeat this trick for each edge adjacent to that vertex and assign pairs of ladder operators to each edge as shown in \autoref{fig:polyhedron-bosons}.

We construct the following operators \cite{Girelli2005Reconstructing,Livine2006Reconstructing} acting on pairs of faces:
\begin{equation}\label{eqn:angle-ops}
	E_{ij} = a^\dag_i a_j + b^\dag_i b_j,
\end{equation}
which satisfy $ E^{\dag}_{ij} = E_{ji} $. Here $ i,j \in \{1,2,\ldots,n\}$ are the face labels, with $ n $ being the total number of faces of the polyhedron. The ladder operators satisfy the commutation relations given in \eqref{eqn:sho-commutator} whenever $ i=j $ and commute otherwise. For $ i=j $, we write these operators as $ E_{ii} = a^{\dag}_i a_i + b^{\dag}_i b_i $. One can check that the operators $ E_{ij} $ satisfy the following commutation relations:
\begin{equation}\label{eqn:u-n-commutators}
	[E_{ij}, E_{kl}] = \delta_{jk} E_{il} - \delta_{il} E_{jk}
\end{equation}
which is the same as the Lie algebra of $ U(n) $. This tells us \cite{Freidel2009The-Fine,Freidel2010UN-Coherent} that the operators $ E_{ij} $ generate a $ \mf{u}(n) $ Lie algebra. One can also verify the following fact:
\begin{equation}\label{eqn:angle-ops-invariance}
	\left[ \sum_{i=1}^n \vec{J}_i, E_{jk} \right] = 0; \quad \forall j,k \in \{1,\ldots,n\}
\end{equation}
where $ \vec{J}_i $ are the angular momentum operators attached to each edge. This tells us that each one of the operators $ E_{ij} $ is invariant under transformations generated by the Gauss constraint \eqref{eqn:gauss-constraint} and therefore define operators which can act on the intertwiner Hilbert space $ \mc{H}_{int} = \text{Inv}_{SU(2)}(\mc{H}_v) $.

The set of operators $ E_{ij} $ constitute only a small subset of all possible operators which can act on the intertwiner Hilbert space. These operators, along-with the $ F_{ij}, F^\dag_{ij} $ operators defined below, are completely analogous to the number, annihilation and creation operators $ a^\dag a, a, a^\dag $ which act on a single harmonic oscillator. The primary difference being that unlike a harmonic oscillator which has an infinite dimensional Hilbert space, the Hilbert space associated with a single face or a collection of faces is finite due to the constraints given in \eqref{eqn:state-correspondence}. 

As shown in \cite{Freidel2009The-Fine} the dimension of the space of intertwiners with $ n $ legs and with total area $ J $ is given by the dimension of the corresponding irreducible representation of $ U(n) $:
\begin{equation}\label{eqn:intertwiner-dimension}
	\dim_n [J] = \frac{1}{J+1} \binom{n+J-1}{J} \binom{n+J-2}{J}
\end{equation}
where $ \binom{n}{k} $ is the binomial coefficient and $ J = \sum_{i=1}^n j_i $ is the total area (using the Freidel-Livine prescription $ A_{FL} $ given in \eqref{eqn:area-operators} for the area of a single face). $ \dim_n [J] $ is, in general much larger than the number of edges. For the special case where $ n = 4 $ and $ j_i = 1/2$ is the same for each face, we find $ \dim_n[J] = 20 $.

Following the discussion in the previous section we know that the diagonal elements $ E_i := E_{ii} $ of the $ \mf{u}(n) $ algebra acting on the state $ \ket{j_i;m_i} $ gives the eigenvalue $ 2 j_i $ which measures \textit{twice} the area of the $ i^\text{th} $ face. Thus $ E = \sum_i E_i $ acting on the intertwiner states has the eigenvalue $ 2 J = 2 \sum_i j_i $ which is \textit{twice} the total area of the polyhedron.

We can also define a set of ladder operators which act on $ \text{Inv}_{SU(2)}(\mc{H}_v) $ as follows:
\begin{equation}\label{eqn:u-n-ladder-ops}
	F_{ij} = a_i b_j - a_j b_i; \quad \forall i \ne j
\end{equation}
which satisfy $ F_{ij} = - F_{ji} $. These operators are also invariant under $ SU(2) $ rotations:
\begin{equation}\label{eqn:ladder-op-invariance}
	\left[ \sum_{i=1}^n \vec{J}_i, F_{jk} \right] = 0,
\end{equation}
and therefore will also act on $ \text{Inv}_{SU(2)}(\mc{H}_v) $.

We will construct (following \cite{Freidel2009The-Fine,Freidel2010UN-Coherent}) $ U(N) $ coherent states in analogy with the construction of the ordinary harmonic oscillator coherent states. Now, recall that in the definition of harmonic oscillator coherent states there is a single complex parameter $ \alpha $ whose real and imaginary parts represent the expectation values of the position and momentum operators, respectively, in that state. In other words, the definition of the coherent state itself encodes the co-ordinates of the point in the classical phase space to which that state corresponds (upto the respective uncertainties in the position and momentum operators). Moreover we can construct the coherent state $ \ket{\alpha} $ by acting upon the vacuum $ \ket{0} $ with the ``displacement operator'' given by:
\begin{equation}\label{eqn:displacement-op}
	D(\alpha) = e^{i(\alpha a - \alpha* a^\dag)},
\end{equation}
such that: $ \ket{\alpha} = D(\alpha) \ket{0} $. Here $ a, a^\dag $ are the usual ladder operators.

One might expect that in order to construct coherent states for intertwiners, in analogy with the harmonic oscillator construction, one could act on some vacuum state $ \ket{0}_{\text{Inv}} \in \text{Inv}_{SU(2)}(\mc{H}_v) $ with a generalized displacement operator constructed as an exponential of a linear combination of the $ U(n) $ ladder operators $ F_{ij}, F^\dag_{ij} $ defined in \eqref{eqn:u-n-ladder-ops}.

The coefficients of the linear combination would be some collection of complex numbers (in analogy with $ \alpha $ for the harmonic oscillator) which would correspond to a point in the \textit{classical phase space}. This expectation does indeed turn to be true. To complete this picture one needs to understand how to label points in the classical phase space associated with an arbitrary graph $ \Gamma $. This can be done via the spinorial formulation of LQG. It also happens to be the point of contact between the physics of intertwiners and that of gluon scattering amplitudes.

\subsection{Spinorial LQG}\label{sec:spinorial-lqg}

The spinorial formulation of LQG was developed in a series of seminal papers by Freidel and Speziale \cite{Freidel2010Twisted,Freidel2010From}. The classical phase space of a single spin network edge is given by $ T^*[SU(2)] \simeq SU(2) \times \mf{su}(2) $ - the cotangent bundle on $ SU(2) $. Configuration variables are group elements $ g \in SU(2)$ and the momenta are elements of the lie algebra $ \mf{su}(2) $. Recall that the Hilbert space of a \textit{single} edge is given by \eqref{eqn:edge-hilbert-space} the space of square integrable functions $ \mc{L}^2[SU(2)] $ on $ SU(2) $. This space can be viewed as the quantization of the classical phase space $ T^*[SU(2)] $.

\begin{figure}
	\centering
	\includegraphics[width=\linewidth]{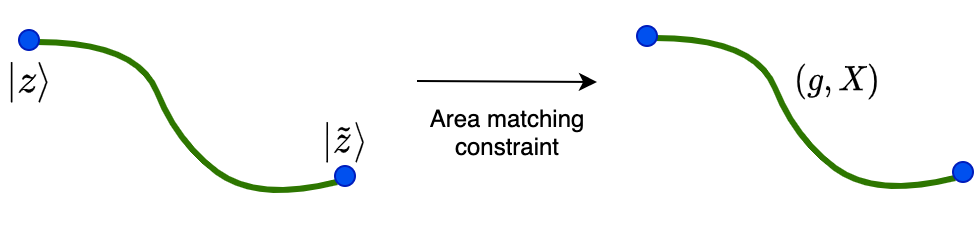}
	\caption{In spinorial LQG we label each end of an edge with spinors $ (\ket{z}, \ket{\tilde z}) \in \C^2 \times \C^2$. Applying the area matching constraint $ \norm{z}^2 - \norm{\tilde z}^2 = 0 $, we obtain the phase space variables $ (g,X) $, with $ g \in SU(2) $ and $ X \in \mf{su}(2) $.}
	\label{fig:spinorial-lqg}
\end{figure}

\begin{figure}[htbp]
	\centering
	\includegraphics[width=0.8\linewidth]{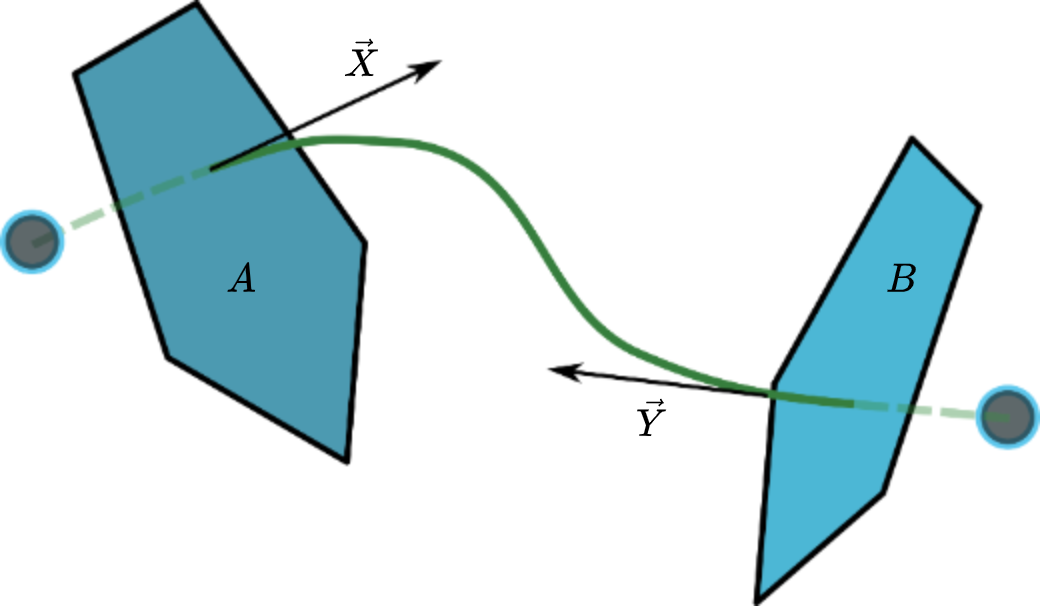}
	\caption{The normal vectors to the faces at each end of an edge in terms of the spinors attached to each end.}
	\label{fig:spinor-normals}
\end{figure}

What was shown in the works by Freidel and Speziale was that one can obtain $ T^*[SU(2)] $ by starting from the space $ \C^2 \times \C^2 $ whose elements are pairs of spinors $ \ket{z}, \ket{\tilde z} $. From each spinor we can construct a three-vector as follows:
\begin{equation}\label{eqn:spinors-to-face-normal}
	\vec{X} = \expectop{z}{\vec{\sigma}}{z}; \quad \vec{\tilde X} = \expectop{\tilde z}{\vec{\sigma}}{\tilde z}
\end{equation}
where $ \vec{\sigma} = (\sigma_x, \sigma_y, \sigma_z)$ are the three Pauli matrices. For a brief introduction to spinorial notation we refer the reader to Appendix \ref{sec:spinor-notation}. These vectors are then to be thought as corresponding to the normals to the faces which are dual to a given edge (\autoref{fig:spinor-normals}). In the classical picture, the area of a face is given by the length of the vector normal to it. Thus, the vectors $ \vec{X}, \vec{\tilde X} $ are the classical counterparts of the angular momentum operators $ \vec{J} $ assigned to each end of an edge. These vectors satisfy the identity:
\begin{equation}\label{eqn:vector-norm}
	\norm{X} = \innerp{z}{z}; \quad \norm{\tilde X} = \innerp{\tilde z}{\tilde z}
\end{equation}
We can construct the momentum variable - the element of $ \mf{su}(2) $ - associated with each end of a vertex by taking the corresponding face normals and contracting them with the Pauli matrices:
\begin{equation}\label{eqn:momentum-variables}
	X = \vec{X} \cdot \vec(\sigma); \quad \tilde X = \vec{\tilde X} \cdot \vec{\sigma}
\end{equation}

We can construct the group element $ g \in SU(2) $ corresponding to the holonomy along the given edge (and plays the role of the configuration variable of the classical phase space $ T^*[SU(2)] $) as follows:
\begin{equation}\label{eqn:spinors-to-holonomy}
	g(z,\tilde z) = \frac{\ket{z}\rbra{\tilde z} - \rket{z}\bra{\tilde z}}{\norm{z} \norm{\tilde z}}
\end{equation}

Now, in order for the spin network to correspond to a triangulation of our background manifold, we must be able to glue the faces labelled $ A $ and $ B $ in \autoref{fig:spinor-normals} to each other. This implies the area of the two faces must match. Since the area of a face is proportional to the length of the face normal, this implies the following \textit{area matching constraint} on the two spinors $ \ket{z} $ and $ \ket{\tilde z} $:
\begin{equation}\label{eqn:area-matching}
	\mc{M} = \innerp{z}{z} - \innerp{\tilde z}{\tilde z} = 0,
\end{equation}
must hold. It is the imposition of this constraint which allows us to reduce the spinorial phase space $ \C^2 \times \C^2 $ to the phase space of a single edge $ T^*[SU(2)] $.

In \cite{Freidel2010From,Freidel2010Twisted} the authors showed the following correspondence:
\begin{equation}\label{eqn:symplectic-reduction}
	T^*[SU(2)] \simeq (\C^2 \times \C^2)//U(1),
\end{equation}
where the ``double quotient'' is understood as taking those elements of $ \C^2 \times \C^2 $ which live on the constraint surface determined by \eqref{eqn:area-matching} and which Poisson commute with this constraint. It is understood that degenerate points, \ie those for which $ \norm{X} = 0 $, $ \innerp{z}{z} = 0 $, $ \innerp{\tilde z}{\tilde z} = 0 $ are excluded from both sides of the above relation.

\begin{figure}[htbp]
	\centering
	\includegraphics[width=0.5\linewidth]{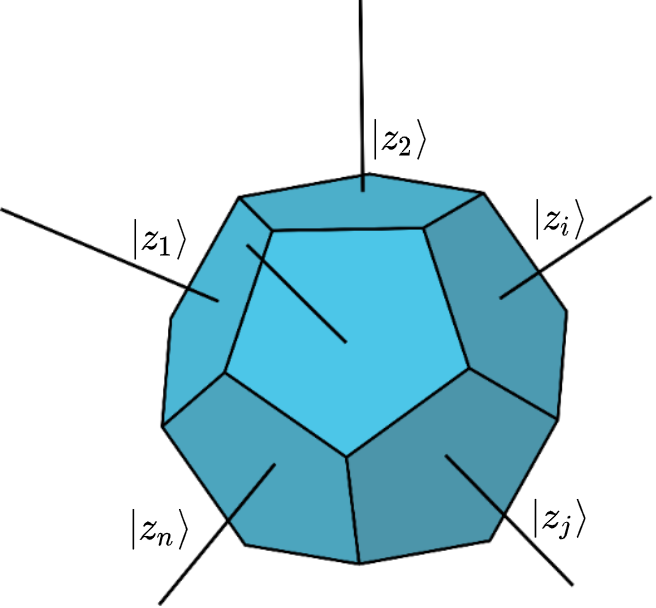}
	\caption{The classical phase space of an intertwiner. Each leg is labelled by a spinor $ \ket{z_i} \in \C^2 $ where $ i \in \{1,\ldots,n\} $.}
	\label{fig:polyhedron-spinors}
\end{figure}

We can now finally address what we set out to do - labelling the classical phase space of the intertwiner. For this purpose we simply need to label the ends of intertwiner legs which meet the given vertex with spinor $ \ket{z_i}; i \in \{1,\ldots,n\} $ as shown in \autoref{fig:polyhedron-spinors}.

It is this set of spinors which provides us with the set of labels needed to parametrize a given $ U(N) $ coherent state.

\subsection{$ U(n) $ Coherent States and the Grassmannian}\label{sec:u-n-coherent-states}

We can now define $ U(n) $ coherent states. These are defined as follows:
\begin{equation}\label{eqn:u-n-coherent-states}
	\bket{J, z_i} = \frac{1}{\sqrt{J+1}}\frac{(F^\dag_\bz)^J}{J!}\bket{0}.
\end{equation}
Here $ \bket{0} $ is the Fock vacuum, which satisfies $ F_{ij} \bket{0} = 0,~\forall i,j $; $ F_{ij} $ is the $ U(n) $ annihilation operator defined in \eqref{eqn:u-n-ladder-ops}; $ J $ measures the total area of the coherent polyhedron represented by this state; and $ \bz \equiv z_{ij} $ is a matrix defined in terms of $ \{z_i\} $ which are the spinor labels assigned to the end of each edge adjacent to the given vertex as shown in \autoref{fig:polyhedron-spinors}. As explained in the previous section, these spinors label a point in the classical phase space of $ n $ sided polyhedra. Using these spinors we define:
\begin{equation}\label{eqn:spinor-matrix}
	\bz \equiv z_{ij} = \innerpB{z_i}{z_j}
\end{equation}
where the inner product in the above expression is defined in \eqref{eqn:spinor-bilinears}. Using the definition of $ z_{ij} $ one can show that $ z_{ij} = - z_{ji} $, i.e. that $ \bz $ is an anti-symmetric matrix. Finally the quantity $ F^\dag_\bz $ in \eqref{eqn:u-n-coherent-states} is defined as follows:
\begin{equation}\label{eqn:ladder-ops-v2}
	F_\bz = \onehalf \sum_{ij} \bar z_{ji} F_{ij}; \quad F^\dag_\bz = \onehalf \sum_{ij} z_{ji} F^\dag_{ij}.
\end{equation}

In the definition of these coherent state \eqref{eqn:u-n-coherent-states} there is a parameter $ J $. One can see that this parameter corresponds to the total area of the $ n $ sided polyhedron as follows. A straightforward computation shows that:
\begin{equation}\label{eqn:e-f-commutator}
	[E, F^\dag_\bz] = 2 F^\dag_\bz
\end{equation}
where $ E = \sum_i E_i $; and $ E_i := E_{ii} $ are the diagonal generators of the $ \mf{u}(n) $ algebra. Using this expression along with standard commutator identities one can show that:
\begin{equation}\label{eqn:e-f-commutator-v2}
	[E, (F^\dag_\bz)^J] = 2 J (F^\dag_\bz)^J; \quad \forall J \in \mathbb{N}.
\end{equation}
Applying this expression to the definition of the coherent states \eqref{eqn:u-n-coherent-states} one can show that:
\begin{equation}\label{eqn:e-eigenvalue}
	E \bket{J, z_i} = 2 J \bket{J, z_i}
\end{equation}
From the discussion following \eqref{eqn:angle-ops-invariance} we know that eigenvalues of the operator $ E $ correspond to \textit{twice} the area of the polyhedron. Therefore we can identify $ J $ in the above expressions as the total area of the polyhedron.

It can be shown explicitly \cite{Freidel2010UN-Coherent} that these states \eqref{eqn:u-n-coherent-states} are indeed generalized coherent states in the sense of Perelomov \cite{Perelomov1977Generalized} and are generated by the action of elements of $ U(n) $ on the highest weight vector. As a result they are also minimum uncertainty states in that the following ratio: $ \Delta E_{ij}/\expect{E_{ij}} \sim 1/\sqrt{J} $ vanishes in the limit of large values for the total area $ A_{tot} = 2 J  $ of the polyhedron. Here:
\begin{equation}\label{eqn:e-op-uncertainty}
	\Delta^2 E_{ij} = \expect{E_{ij}E_{ji}} - \expect{E_{ij}} \expect{E_{ji}}
\end{equation}
is the dispersion in the expectation value of the operators $ E_{ij} $. Having shown that \eqref{eqn:u-n-coherent-states} satisfy all the properties expected of coherent states representing semiclassical states of geometry, we can now finally come to the central fact - that these states are labeled by elements of the Grassmannian $ Gr_{2,n} $. We can see this as follows.

These states $ \bket{J, z_i} $ have the following important property under the diagonal action of elements $ \lambda \in GL(2,\C) $ on the spinors $ \{z_i\} $:
\begin{equation}\label{eqn:gl2c-action}
	\bket{J, \lambda z_i} = (\det \lambda)^J \bket{J,z_i}
\end{equation}
Under the diagonal action of $ GL(2,\C) $, each of the spinors transforms as $ \ket{z_i} \rightarrow \lambda \ket{z_i} $. Using direct computation one can confirm that the matrix $ \bz $ transforms as $ \bz \rightarrow (\det \lambda) \bz $ under this transformation. So does the ladder operator $ F^\dag_\bz \rightarrow (\det \lambda)  F^\dag_\bz $. Finally we have $ (F^\dag_\bz)^J \rightarrow (\det \lambda)^J  (F^\dag_\bz)^J $.

Now these Freidel-Livine coherent states are labeled by $ 2n $ complex numbers given by the set of $ n $ spinors $ \{z_i\}; i \in 1, \ldots, n $. In other words they are labeled by points in $ \C^{2n} $. In addition under the action of $ GL(2,\C) $ these states transform as homogeneous functions of degree $ J $ as shown in \eqref{eqn:gl2c-action}. Thus we can make the following assertion:

\begin{quote}
	\textit{The space of unique labels of the FL coherent states is given by $ C^{2n}/GL(2,\C) $ which is precisely the Grassmannian $ Gr_{2,n} $ - the set of all $ 2 $ planes in $ C^n $.}
\end{quote}

\subsection{Reality Condition on Momenta}\label{sec:reality-condition}

At this point the natural question to ask is what is the relation between the momenta of $ n $ massless particles as defined in \eqref{eqn:null-momenta} and the spinors labeling coherent intertwiners. Recall that in terms of a pair of spinors $ \lambda_\alpha, \tilde \lambda_{\dot \alpha} $ a momentum 4-vector can be expressed as:
\begin{equation}\label{eqn:spinor-to-vector-1}
	p_{\alpha \dot \alpha} = \lambda_\alpha \tilde \lambda_{\dot \alpha}	
\end{equation}
The requirement that $ p_{\alpha \dot \alpha} $ represent a real momentum reduces to the statement that:
\begin{equation}\label{eqn:spinor-reality}
	\tilde \lambda_{\dot \alpha} = \pm (\lambda^*_\alpha)	
\end{equation}
We choose the positive sign in this expression in order to select positive energy momenta.

Given the spinor $ \ket{z} $ we can construct a future-pointing null 4-vector $ X^\mu = (X^0, X^i)$ whose components are given by \cite{Freidel2010From}:
\begin{equation}\label{eqn:spinor-to-vector-2}
	X^0 = \onehalf \innerp{z}{z}; \quad X^i = \expectop{z}{\vec \sigma}{z}
\end{equation}
The converse of the above equation becomes:
\begin{equation}\label{eqn:vector-to-spinor-1}
	\ket{z}\bra{z} = X^0 \id + X^i \sigma_i
\end{equation}
We can rewrite the above expression as:
\begin{equation}\label{eqn:spinor-to-vector-3}
	X_{\alpha \dot \alpha} = z_\alpha z^*_{\dot \alpha} \equiv (\ket{z}\bra{z})_{\alpha \dot \alpha}
\end{equation}
It is easy to see that the expression \eqref{eqn:spinor-to-vector-1} can be expressed in precisely the same manner due to the reality condition \eqref{eqn:spinor-reality}, as:
\begin{equation}\label{eqn:spinor-to-vector-4}
	p_{\alpha \dot \alpha} = \lambda_\alpha \lambda^*_{\dot \alpha} = (\ket{\lambda}\bra{\lambda})_{\alpha \dot \alpha}
\end{equation}
Thus the spinors $ \ket{z} $ used to label coherent intertwiners and the spinors $ \ket{\lambda} $ used to label massless particle momenta can be viewed as one and the same thing. To each edge of an intertwiner we can therefore assign a \textit{real}, null momentum 4-vector with components given by \eqref{eqn:spinor-to-vector-2} and in our interpretation this null vector is to be understood as representing the momentum of a massless particle.

\section{Discussion: Quantum Gravity on the Positive Grassmannian}\label{sec:discussion}

In this work we have pointed out an interesting mathematical coincidence. The kinematic space of states for the scattering of $ n $ massless particles is identical to the space of states of semiclassical states of quantum geometry represented by Freidel-Livine coherent intertwiners.

It would be easy to dismiss this correspondence simply as a mathematical accident, were it not for the fact that it provides us with a compelling physical picture for the emergence of classical spacetime purely from considering scattering amplitudes of $ n $ massless particles, in line with the expectations of the Amplituhedron program \cite{Arkani-Hamed2014The-Amplituhedron}. Conversely it suggests that we can view Freidel-Livine coherent intertwiners as representing the scattering of $ n $ massless particles in the framework of a background independent theory of quantum gravity.

We wish to stress that much of this paper simply recounts the result of previous works. However, to our knowledge, we are the first to recognize the significance of the correspondence between the two seemingly disparate topics of scattering amplitudes and coherent intertwiners. This correspondence has several important physical implications which we spell out in the following sections.



\subsection{Particle Momenta, Area and the Classical Limit}\label{sec:momenta-area}

We would like to point out a conundrum which results from taking such a correspondence to be valid. The norm $ \innerp{z_i}{z_i} $ of a spinor $ \ket{z_i} $ attached to a given edge gives us the length of the vector $ \vec X_i $ normal to the face punctured by the given edge. This length is the classical counterpart of the angular momentum operator $ \vec{J} $ which measures the area of the face. Therefore $ \innerp{z_i}{z_i} $ is a measure of the face area. Now, in the context of particle scattering, $ \vec{X_i} $ is the spatial part of the null four vector representing the particle's momentum and its length is a measure of the energy (or momentum) of that particle.

What this tells us is that small particle energies correspond to small areas whereas large energies correspond to polyhedra with large face areas. On the other hand, as we discussed previously, the semiclassical limit of the intertwiner state corresponds to the limit in which the polyhedron area $ 2J $ becomes large. These two observations then imply that the semiclassical limit of the intertwiner state corresponds to the scattering of high energy particles, whereas the deep quantum part - when the values of face areas are close to the limit $ l_p^2 $ - corresponds to the scattering of ``soft'' (low energy) particles. This is counter to the common intuition according to which the momentum of a particle is inversely proportional to the length scales probed by that particle.

If we take this correspondence at face value it further suggests that the momenta of massless particles is bounded from \textit{below} by the minimum possible quantum of area which is of the order of $ l_p^2 $. This would imply that quantum gravity plays the role of an infrared regulator. While we do not have a complete physical picture at present, we can present a plausible argument based on the $ \textbf{Momentum} = \textbf{Complexity} $ correspondence for this counter-intuitive suggestion.

\subsection{Kinematical vs Dynamical Aspects}\label{sec:kin-vs-dyn}

\subsection{Computational Complexity of Scattering Processes}\label{sec:comlexity}

It is no longer a matter of wild speculation to suggest that in a theory of quantum gravity, spacetime geometry, and all processes taking place ``on'' that geometry, should have a description in terms of a quantum circuit \cite{Czech2018Einstein,Caputa2018Quantum,Takayanagi2018Holographic,Camargo2019Path}. The natural picture which emerges from these considerations is that one should think of any scattering process as a computational process where the ``in'' and ``out'' states should be viewed as the ``input'' and ``output'' states of some quantum circuit. Now one can ask the following question:
\begin{quote}
	\textit{What is the computational complexity of a given scattering process?}
\end{quote}
We can imagine the underlying quantum circuit as consisting of some sequence of elementary gates which act on the input ``wires'' in a pair-wise manner.

There are two measures of the complexity of a circuit - the circuit width and the circuit depth. The circuit width is determined by the total number of ``wires'' involved in the associated scattering process. For instance if we consider $ 2\rightarrow 2 $ scattering then the width of circuit will be $ 2 $. The circuit depth refers to the number of layers of gates which separate the output states from the input states. The complexity measure we are interested in is the circuit depth.

Now, an intertwiner is basically a quantum gate - it provides a unitary transformation which acts on some set of ``incoming'' edges and gives the result in terms of a set of ``outgoing'' edges. If we consider the Fock vacuum $ \bket{0} $ as our minimum complexity reference state, then we can see from the form of the intertwiner coherent states \eqref{eqn:u-n-coherent-states} that as the area value $ 2J $ associated with such a state increases so does the ``complexity'' (in terms of the number of operations required to generate the state starting from the Fock vacuum $ \bket{0} $).

We can now use a particular expression of quantum circuit picture of spacetime, in the form of the $ \textbf{Momentum} = \textbf{Complexity} $ correspondence studied by Barb\'on and collaborators \cite{Barbon2019Momentum/Complexity,Barbon2020Proof} to explain why the magnitude of particle momenta should scale as the area of FL intertwiners, counter to what intuition suggests. What has been shown in these works is that the complexity of a given region of spacetime changes if momentum either falls into or comes out of that region. Very roughly speaking the change in the complexity $ \delta C $ of some bounded region of spacetime $ S $ is proportional to the total momentum $ \delta P $ which enters (or leaves) that region. For simplicity we can take $ S $ to be a sphere. Only the radial (transverse to the surface) components of the momentum contribute to this change.

Now consider $ 2 \rightarrow 2 $ scattering of massless particles. As two particles with total (radial) momentum $ P_{in} = p_1 + p_2 $ enter the region the complexity of the state of that region $ \ket{\Psi_S} $ increases by an amount $ \delta C \propto P_{in} $. When the outgoing particles with momentum $ P_{out} = P_{in} $ leave the region the complexity reduces by the same amount. This tells us that as the total incoming momentum increases the circuit complexity, measured in terms of the number of gates required to generate the state of region $ S $, also increases. This is in concordance with the observation made earlier that the complexity of a FL intertwiner increases with increasing area. Hence, this picture justifies associating small momenta with ``less'' complex intertwiner states with small areas and large momenta with ``more'' complex intertwiner states with large areas.

\subsection{A Theory of Quantum Gravity on the Grassmanian}\label{sec:qgtheory}

The considerations in the previous sections allows us to outline what a possible theory of quantum gravity, where the basic ingredients are scattering amplitudes, might look like. We have shown that there is a one-to-one correspondence between scattering amplitude of $ n $ massless particles and a single $ n $-valent spin-network vertex. The question now arises whether we can understand how the two sides of the equation: \textbf{matter $ \equiv $ gravity} arise in this picture. Can the dynamics of $ n $ massless particles have a physical effect on the gravity sector or vice-versa? This is indeed possible in the following manner.

Consider two copies of the Grassmannian manifold $ Gr_{2,n} $ which we will label as $ \mc M_{grav} $ and $ \mc M_{matter} $. Consider a point $ p \in \mc M_{grav} $ and a copy of the same point $ q \in \mc M_{matter} $. The point $ p $ corresponds to kinematic data for $ n $ particle scattering and $ q $ represents the geometric data for a $ n $-valent intertwiner.

We would ordinarily impose momentum conservation on any scattering process so the point $ p $ would represent kinematic data of $ n $ particles both, \textit{before} and \textit{after} the scattering event. Now, imagine a situation where the total momentum in a scattering process is \textit{not conserved}. In this case the point $ q \in \mc M_{matter} $ would represent the kinematic data \textit{before} the scattering event and a \textit{different} point $ q' = q - \delta q \in \mc M_{matter} $ would represent the kinematic data \textit{after} the scattering. Clearly, if this was the whole picture we would be in trouble, because in any sensible theory momentum must be conserved in any physical process! However, the existence of the gravitational sector, given by the point $ p \in \mc M_{grav} $ on the second copy of the Grasmannian manifold gives us a way out of this dilemma.

The total momentum lost in the scattering event in going from the point $ q \rightarrow q' = q - \delta q $ can be compensated by changing the point $ p $ to $ p + \delta p $. What this would imply physically is that the net loss of momentum in the scattering event has been transferred to the gravitational degrees of freedom of the $ n $-valent intertwiner. Geometrically this would imply that the area of the classical polyhedron represented by the coherent intertwiner increases by an amount sufficient to compensate for the loss of momentum in the scattering event. In this way we have provided a tentative outline of the fundamental equation: \textbf{matter $ \equiv $ gravity} in the language of the dynamics of point particles living on two copies of the Grassmannian manifold $ Gr_{2,n} $. Further details of this picture are under investigation by the authors.

\begin{acknowledgements}
	The authors would like to thank Pritam Nanda for helpful discussions. DV would like to thank his colleagues in the Physics department at NITK Surathkal for supporting his research efforts. DV thanks the Inter-University Center for Astronomy and Astrophysics (IUCAA), Pune for support under the Visiting Associates program. The authors would also like to thank their respective families for continuing support and inspiration.
\end{acknowledgements}

\appendix

\section{Spinor Notation}\label{sec:spinor-notation}

In this section we present some basic notations and relations involving spinors \cite{Livine2011Loop} which are required for understanding some of the expressions in the main text. A spinor $ \ket{z} $ is an element of $ \C^2 $, while $ \bra{z} $ is its hermitian adjoint:
\begin{equation}\label{eqn:spinors}
	\ket{z} = \begin{pmatrix}
		z^0 \\ z^1
	\end{pmatrix}; \quad \bra{z} = \begin{pmatrix}
									\bar z^0 & \bar z^1
									\end{pmatrix}
\end{equation}
The ``dual spinor'' $ \rket{z} $ is defined in terms of the spinor $ \ket{z} $ as follows:
\begin{equation}\label{eqn:dual-spinors}
	\rket{z} = \epsilon \rket{\bar z} = \begin{pmatrix}
											-\bar z^1 \\ \bar z^0
										\end{pmatrix}
\end{equation}
where $ \epsilon $ is the antisymmetric tensor with components:
\begin{equation}\label{eqn:epsilon-tensor}
	\epsilon = \begin{pmatrix}
					0 & -1 \\
					1 & 0
				\end{pmatrix}
\end{equation}
The combination of complex conjugation $ \ket{z} \rightarrow K \ket{z} = \ket{\bar z} $ followed by multiplication by $ \epsilon $ can be written as the map: $ J = \epsilon K $ which satisfies $ J^2 = -\id $. There are two different spinor bilinears:
\begin{align}\label{eqn:spinor-bilinears}
	\innerp{x}{y} & = \rinnerp{y}{x} = \bar x^0 y^0 + \bar x^1 y^1; \nonumber \\
	\innerpB{x}{y} & = -\innerpB{y}{x} = x^0 y^1 - x^1 y^0
\end{align}
From any spinor $ \ket{z} $ one can construct a Hermitian matrix which can be expanded in terms of the $ (\id, \vec{\sigma}) $ basis as follows:
\begin{equation}\label{eqn:spinor-matrix}
	\ket{z}\bra{z} = \onehalf (\innerp{z}{z} \id + \vec{X} \cdot \vec{\sigma})
\end{equation}
where $ \vec{X} \in \mbb{R}^3$ and $ \norm{X} = \innerp{z}{z}$. The above map is not one-to-one because a $ U(1) $ rotation $ \ket{z} \rightarrow e^{i\theta} \ket{z}$ leaves the matrix $ \ket{z}\bra{z} $, and hence the vector $ \vec{X} $, invariant. As mentioned in the text we can construct an element $ g \in SU(2) $ from two spinors via:
\begin{equation}
	g(z,\tilde z) = \frac{\ket{z}\rbra{\tilde z} - \rket{z}\bra{\tilde z}}{\norm{z} \norm{\tilde z}}
\end{equation}
The inverse of this element is equal to its Hermitian adjoint by unitarity of $ SU(2) $ and is given by:
\begin{equation}\label{eqn:holonomy-inverse}
	g^{-1}(z, \tilde z) = g^\dag(z, \tilde z) = \frac{\rket{\tilde z}\bra{z} - \ket{\tilde z}\rbra{z}}{\norm{z} \norm{\tilde z}}
\end{equation}
It is easy to verify using the spinor bilinear identities given in \eqref{eqn:spinor-bilinears} that indeed: $ g^{-1} g = \id $. It takes a little more work to verify that $ \det(g) = 1 $. The action of the holonomy on spinors (when the area matching constraint \eqref{eqn:area-matching} is satisfied) is given by:
\begin{equation}\label{eqn:holonomy-action}
	g \frac{\ket{\tilde z}}{\norm{\tilde z}} = -\frac{\rket{z}}{\norm{z}}; \quad g^{-1} \frac{\ket{z}}{\norm{z}} = \frac{\rket{\tilde z}}{\norm{\tilde z}}
\end{equation}
Using \eqref{eqn:spinors-to-holonomy} one can verify that under the \textit{local} action of $ SU(2) $ on a spinor pair:
\begin{equation}\label{eqn:su2-spinor-action}
	(\ket{z}, \ket{\tilde z}) \rightarrow (h_1 \ket{z}, h_2 \ket{\tilde z}),
\end{equation}
the holonomy $ g(z, \tilde z) $ transforms exactly as a holonomy should:
\begin{equation}\label{eqn:su2-holonomy-action}
	g(z, \tilde z) \rightarrow h_1 g(z, \tilde z) h_2^{-1},
\end{equation}
where $ h_1, h_2 \in SU(2) $ are gauge transformations acting on the source and target vertices of a given edge respectively. One can verify that when \eqref{eqn:area-matching} holds, the $ \mf{su}(2) $ element $ X = \vec{X}\cdot\vec{\sigma} $ (corresponding to $ \ket{z}) $ and the element $ \tilde{X} = \vec{\tilde X}\cdot\vec{\sigma} $ (corresponding to $ \ket{\tilde z} $) satisfy:
\begin{equation}\label{eqn:3-vecs-relation}
	\tilde X = -g^{-1} X g
\end{equation}

\bibliographystyle{JHEP3}


\bibliography{lqg-amplituhedron.bib}


\end{document}